\documentclass[11pt]{article}
\usepackage[margin=1in]{geometry}

\newtheorem{theorem}{Theorem}
\newtheorem{definition}{Definition}

\newtheorem{proposition}{Proposition}

\usepackage{amsfonts}
\usepackage[fleqn]{amsmath}
\usepackage{bbm}
\usepackage{float}
\usepackage{caption}
\usepackage{subcaption}
\usepackage{xcolor}
\usepackage{mathtools}
\usepackage{tikz}

\usetikzlibrary{shapes.geometric,arrows}
\usepackage{multicol}
\usepackage{stackrel}
\renewcommand{\)}{\right)}
\renewcommand{\(}{\left(}
\usepackage{subcaption}
\usepackage{amssymb}
\usepackage{graphicx}      
  \usepackage{natbib}
\begin{document}

\title{Learning an Unknown Network State in Routing Games
}


\author{Manxi Wu, and Saurabh Amin
\thanks{M. Wu is with the Institute for Data, Systems, and Society, S. Amin is with the Department of Civil and Environmental Engineering, Massachusetts Institute of Technology (MIT), Cambridge, MA, USA
        {\tt\small \{manxiwu,amins\}@mit.edu}}}%
\date{}
\maketitle

\begin{abstract}                
We study learning dynamics induced by myopic travelers who repeatedly play a routing game on a transportation network with an unknown state. The state impacts cost functions of one or more edges of the network. In each stage, travelers choose their routes according to Wardrop equilibrium based on public belief of the state. This belief is broadcasted by an information system that observes the edge loads and realized costs on the used edges, and performs a Bayesian update to the prior stage's belief. We show that the sequence of public beliefs and edge load vectors generated by the repeated play converge almost surely. In any rest point, travelers have no incentive to deviate from the chosen routes and accurately learn the true costs on the used edges. However, the costs on edges that are not used may not be accurately learned. Thus, \emph{learning can be incomplete} in that the edge load vectors at rest point and complete information equilibrium can be different. We present some conditions for complete learning and illustrate situations when such an outcome is not guaranteed.
\end{abstract}

\textbf{keywords:}
Multi-agent systems, Decentralized, distributed, and cooperative estimation, Strategic learning, Routing games, Bayesian state estimation.


\newcommand{\I}{I}
\newcommand{\lambi}{\lambda^i}
\newcommand{\lamb}{\lambda}
\renewcommand{\i}{i}
\newcommand{\R}{\mathcal{R}}
\newcommand{\E}{\mathcal{E}}
\newcommand{\e}{e}
\newcommand{\Chij}{X^j}
\newcommand{\Chibarj}{\bar{X}^{j}}
\newcommand{\pst}{p(\type|\s)}
\newcommand{\chij}{x^j}
\newcommand{\chibarj}{\bar{x}^{j}}
\newcommand{\chiseq}{x}
\newcommand{\chibarseq}{\bar{x}}
\renewcommand{\(}{\left(}
\renewcommand{\a}{a}
\newcommand{\n}{n}

\newcommand{\wsran}{w^{\sran*}}
\renewcommand{\)}{\right)}
\newcommand{\deleq}{\stackrel{\Delta}{=}}
\newcommand{\Sbar}{\bar{\S}(\wbar)}
\newcommand{\q}{q}
\renewcommand{\r}{r}
\newcommand{\qsran}{q^{\sran}}
\newcommand{\whist}{W^{\t}}
\newcommand{\etabar}{\bar{\eta}(\s)}
\newcommand{\whis}{W}
\newcommand{\Q}{Q}
\newcommand{\ti}{t^i}
\newcommand{\typeran}{\mathbf{\type}}
\newcommand{\Ti}{\T^{\i}}
\newcommand{\type}{t}
\renewcommand{\S}{\mathcal{S}}
\newcommand{\D}{D}
\newcommand{\etaj}{\eta^j(\s)}
\newcommand{\qti}{q(\ti)}
\newcommand{\qrti}{q_\r(\ti)}
\newcommand{\ce}{c_\e}
\newcommand{\epes}{\epsilon_\e^{\s}}
\newcommand{\eps}{\epsilon^\s}
\newcommand{\qt}{q^\t}
\newcommand{\qrt}{q^\t_\r}
\newcommand{\tiran}{\mathbf{\ti}}
\newcommand{\wet}{w_e^\t}
\newcommand{\wt}{w^\t}
\newcommand{\ept}{\epsilon^{\t}}
\newcommand{\epet}{\ept_e}
\newcommand{\epesran}{\epsilon_\e^{\sran}}
\newcommand{\pstype}{p^{\sran,\typeran}}
\newcommand{\qbar}{\bar{\q}}
\newcommand{\ct}{c^\t}
\newcommand{\cet}{c_e^\t}
\newcommand{\Et}{\E^\t}
\newcommand{\Ej}{\E^j}
\newcommand{\wbar}{\bar{w}}
\newcommand{\webar}{\wbar_e}
\newcommand{\cbar}{\bar{c}}
\newcommand{\Ebar}{\bar{\E}}
\newcommand{\thetabar}{\bar{\theta}}
\newcommand{\w}{w}
\renewcommand{\wp}{w'}

\newcommand{\wwet}{w^{t*}}
\newcommand{\playpath}{C}
\newcommand{\Playpath}{\mathcal{H}}
\newcommand{\Fplaypath}{\mathcal{F}}
\newcommand{\playpatht}{C^\t}
\newcommand{\Ht}{\Playpath^\t}
\newcommand{\Ft}{\mathcal{F}^\t}
\newcommand{\wtwe}{w^{\t*}}
\newcommand{\qwert}{q^{*}_\r}
\renewcommand{\P}{\mathcal{P}}
\newcommand{\Qt}{\Q^\t}
\newcommand{\ep}{e^{'}}
\newcommand{\thetainfty}{\theta^{\infty}}

\newcommand{\qwes}{q^{*}(\s)}
\newcommand{\Cinfty}{C_{\thetainfty}}
\newcommand{\qwet}{q^{\t*}}
\newcommand{\wwe}{w^{*}}
\newcommand{\ect}{\mathbb{E}_{\thetat}[c^\s(\wwe)]}
\newcommand{\wwebart}{\bar{w}^{*t}}
\newcommand{\wewebart}{\bar{w}_\e^{*t}}
\newcommand{\csrant}{\ell^\sran (\wwe)}
\newcommand{\Shat}{\widehat{S}}
\newcommand{\shat}{\hat{\s}}
\newcommand{\crsran}{\ell_\r^\sran}
\newcommand{\crsrant}{\ell_r^\sran (\qtwe)}
\newcommand{\crsranthat}{\ell_r^\sran (q^{\that*}(\thetathat))}
\newcommand{\ecwett}{\mathbb{E}_{\thetat}\left[c^s(\qwet)\right]}
\newcommand{\ecrwett}{\mathbb{E}_{\thetat}\left[c_r^s(\q^{\t*} )\right]}
\newcommand{\ecweinftyt}{\mathbb{E}_{\thetainfty}\left[c^s(\qwet)\right]}
\newcommand{\ecrweinftyt}{\mathbb{E}_{\thetainfty}\left[c_\r^s(\qwet)\right]}
\newcommand{\ecrqbarweinfty}{\mathbb{E}_{\thetainfty}\left[c_\r^\s(\qwebar)\right]}
\newcommand{\ecweinftyinfty}{\mathbb{E}_{\thetainfty}\left[c^s(\qwe(\thetainfty))\right]}
\newcommand{\ecrweinftyinfty}{\mathbb{E}_{\thetainfty}\left[c_\r^s(\qwebar)\right]}
\newcommand{\wewet}{w_\e^{\t*}}
\newcommand{\wewethat}{w_\e^{\that*}}
\newcommand{\thetazero}{\theta^0}

\newcommand{\qr}{\q_\r}
\newcommand{\qwe}{q^{*}}
\newcommand{\Sigt}{\Sigma^\t}
\newcommand{\qrwe}{\q_\r^{*}}
\newcommand{\qrtwe}{\q_\r^{\t*}}
\newcommand{\qweinfty}{q^{\infty*}}
\renewcommand{\t}{k}
\newcommand{\crt}{c_r^\t}
\newcommand{\crtilt}{c^{s}_r}
\newcommand{\crtiltran}{\mathbf{c^{t}_r}}
\newcommand{\conetilt}{c^{t}_1}
\newcommand{\ctwotilt}{c^{t}_2}
\newcommand{\ctilt}{c^\t}
\newcommand{\dist}{\phi[\sran, \wt](\ct)}
\newcommand{\Cs}{C_\sran}

\newcommand{\wetran}{w^{\t}_\e(\typeran)}
\newcommand{\ctran}{\mathbf{\ct}}
\newcommand{\cetran}{\mathbf{\ct_e}}
\newcommand{\distp}{\phi[s', \wt](\ct)}
\newcommand{\ecrt}{\mathbb{E}_{\thetat}[\ell_\r^\s(q^\t)]}
\newcommand{\ecrdagt}{\mathbb{E}_{\thetat}[c_{\rdag}^\s(q^\t)]}
\newcommand{\ecrptwe}{\mathbb{E}_{\thetat}[c_{\r^{'}}^\s(q^{t*})]}
\newcommand{\Ps}{\P^\s}
\newcommand{\Pst}{\P^{\sran, \typeran}}
\newcommand{\ps}{p^\s}
\newcommand{\cets}{\ell_{\Et}^\s}
\newcommand{\ecrqtwe}{\mathbb{E}_{\thetat}[\ell_{\r}^\s(q^{\t*})]}
\newcommand{\ecrqtiwe}{\mathbb{E}^{\t}[\ell_{\r}^\s(q^{\t*})|\ti]}
\newcommand{\ecrpqtwe}{\mathbb{E}_{\thetat}[\ell_{\r'}^\s(q^{\t*})]}
\newcommand{\lambl}{\underline{\lambda}}
\newcommand{\crs}{\ell_\r^\s}
\newcommand{\Einfty}{\E^{\infty}}
\renewcommand{\c}{c}
\newcommand{\belieft}{\beta^{\t}}
\newcommand{\ecrqt}{\mathbb{E}_{\thetat}[\ell_\r^s(q^\t)]}
\newcommand{\ecrqti}{\mathbb{E}_{\theta}[\ell_\r^s(q)|\ti]}
\newcommand{\game}{G}
\newcommand{\gamet}{G^\t}
\newcommand{\T}{T}
\newcommand{\cpt}{\pi^{\t}}
\newcommand{\tmi}{\type^{-\i}}

\newcommand{\Tmi}{\T^{-\i}}
\newcommand{\cesran}{\ell_e^{\sran}}
\newcommand{\thetap}{\theta'}
\newcommand{\qwep}{\q^{*'}}
\newcommand{\ecrqwe}{\mathbb{E}_{\theta}[\ell_\r^\s(\qwe)]}
\newcommand{\ecrqwep}{\mathbb{E}_{\theta'}[\ell_\r^\s(\qwep)]}
\newcommand{\wwep}{w^{*'}}
\newcommand{\thetat}{\theta^\t}
\newcommand{\Sinfty}{\S^{\infty}}
\newcommand{\lambs}{\lambda^s}
\newcommand{\wweinfty}{w^{\infty*}}
\newcommand{\sran}{\mathbf{\s}}
\newcommand{\pro}{\mathrm{Pr}}
\newcommand{\playpathwe}{\playpath^{*}}

\newcommand{\wewe}{w_e^{*}}
\newcommand{\we}{w_e}
\newcommand{\ces}{\ell_\e^\s}
\newcommand{\qwebar}{\bar{q}^{*}}

\newcommand{\distnoise}{\psi^{\s,\t}}
\newcommand{\wwebar}{\bar{w}^{*}}
\newcommand{\wewebar}{\wwebar_e}

\newcommand{\rp}{r^{'}}
\newcommand{\qtwe}{q^{\t*}}
\newcommand{\s}{s}

\section{Introduction}\label{sec:intro}
Transportation networks are prone to disruptions resulting from random infrastructure breakdowns and adverse events  such as natural disasters and security attacks. The impact of these disruptions can be modeled as a sudden change in the latent condition (or state) that influences travel costs on one or more network edges. Major disruptions, such as the 2007 collapse of I-35W bridge over Mississippi River in Minneapolis, can result in an abrupt interruption of nominal flow patterns on the network and trigger repeated learning and adjustment of travel decisions over a period of time (\citet{zhu2010traffic}). In this paper, we study learning dynamics of travelers who make strategic route choices under imperfect state information.  

We focus on learning of the unknown network state when travelers have access to an information system that repeatedly updates and broadcasts the state information (i.e., public belief). This problem is relevant to situations when a disruption or replacement of an infrastructure facility (e.g., bridge, bypass highway) results in travelers' reliance on a public source of information to learn and adjust their route choices. In recent years, such information sources have become widely available and increasingly sophisticated in terms of their ability to collect traffic loads and costs on network edges via a variety of heterogenous data sources (e.g., fixed sensors, GPS-enabled mobile sensors, and crowdsourcing services). Our generic learning model considers strategic travelers with access to imperfect information of the state from a public information system.    

In particular, we model the learning dynamics as a routing game of non-atomic travelers who repeatedly use the network with an unknown state (Sec. \ref{model}). The state represents the latent network condition and affects the travel costs on edges. In each state, the cost of any edge is an increasing function of the edge load (i.e., the aggregate traffic load on that edge). For simplicity, we assume that travelers have identical preferences and that the demand is fixed. At the beginning of each stage, all travelers have access to the most recent public belief of the state (i.e., the probability of each possible state) through the public information system. Travelers then myopically choose routes with the smallest expected cost based on the public belief of the state. Thus, the routing strategy in each stage is a Wardrop equilibrium corresponding to the current public belief. Using classical results (\citet{rosenthal1973class, sandholm2001potential}) one can conclude the essential uniqueness of the induced equilibrium load. Furthermore, the realized edge costs are random functions of the edge load and the true state. Again, for simplicity, we assume that cost distributions are normally distributed. A key feature of our model is that the public information system act as an aggregator: In each stage, it collects the loads and realized costs of all edges that were used in that stage, and updates the belief of state according to Bayes' rule. However, the costs of edges that are not used are not available to the information system.

%


Our work contributes to the literature on learning and information effects in routing games. Prior work has studied the impact of heterogeneous state information on travel decisions in static settings~\citet{arnott1991does, khan2018bottleneck, wu2018value}. On one hand, the transportation community has studied travelers' response to real-time traffic information using dynamic behavioral models and network simulation (\citet{ben1991dynamic, mahmassani1991system}). In a related work \citet{jha1998perception}, the authors propose a Bayesian model to capture how travelers update their perception of travel times based on the received information. Their simulation-based approach identifies the evolution and stabilization of travel patterns. 

On the other hand, a variety of learning dynamics (\citet{fudenberg1998theory}) have been particularized to routing games. In the classical fictitious play setting, players best respond to the belief of opponents' strategies in each stage and revise their beliefs based on the observed actions (\citet{monderer1996potential, marden2009joint}). Another well-known setting is pay-off based learning (\citet{marden2012revisiting, cominetti2010payoff}), in which players' strategies follow a controlled dynamics based on the history of their own realized costs. In contrast, our learning setup focuses on the strategic aspect of travelers' behavior by assuming that they choose routes with the lowest cost based on the belief of state, which repeatedly gets updated by the public information system.  Also our analysis focuses on learning of the uncertain state instead of learning opponents' strategies.

In a related paper~\citet{meigs2017learning}, the authors consider a repeated routing game, in which travelers learn the slope of linear cost functions using a least square estimator based on the realized costs. In this paper, we take a Bayesian viewpoint and do not impose a specific functional form on the edge costs. Importantly, the Bayesian update on state belief is based on the loads and the realized costs on edges that are used by the travelers. Indeed, in our model, history of the play affects the set of used edges and the realized costs through the belief update, and hence the outcomes in future stages. Thus, the stage game outcomes are correlated and non-identical across stages and classical theory in Bayesian learning (for e.g., \citet{blackwell1962merging}) cannot be applied to our problem. 

We now summarize our main result (Theorem \ref{theorem_convergence}) and analysis approach. We show that the public belief and the equilibrium edge load in each stage almost surely converge to a rest point that satisfies two properties: (i) The edge load is induced by a Wardrop equilibrium that corresponds to the public belief at convergence; (ii) Travelers accurately estimate the expected costs of all edges that are used. These two properties ensure that, at rest point, the realized edge costs do not provide new information about the state, and no traveler has an incentive to change her route choice (Sec. \ref{sec_self_confirm}). 

Notably, the concept of rest point is similar to the self-confirming equilibrium studied in \citet{fudenberg1993self}, \citet{fudenberg1995learning}. These papers consider a strategic learning model, in which players myopically play an extensive-form game in each stage, and update their belief of opponents' strategies based on the observed actions. Eventually, players accurately predict and best respond to opponents' strategies on the reached information sets, but may continue to maintain incorrect beliefs on the unreached ones. Similarly, in our model, travelers learn the true cost of the used edges at rest points, but may persistently have an incorrect estimate of costs for the unused edges because no additional information is available to revise the belief. Consequently, some edges that should be taken in the complete information equilibrium may not be taken at a rest point. 

Finally, we study the some properties of rest points. We find that if the network is series-parallel, the average cost experienced by travelers at any rest point is no less than that in complete information equilibrium (Proposition \ref{higher_cost}). We also provide a set of conditions, under each of which travelers end up only using the edges that are part of the complete information equilibrium; i.e., they eventually make route choice as if they know the true state (Proposition \ref{nash_proposition}). In this case, the learning is complete in that the edge load vector converges almost surely to the complete information equilibrium (Sec.\ref{sec:rest_point}).

We provide some concluding remarks in Sec. \ref{sec:conclude}.

\section{Model}\label{model}
In this section, we introduce our model of learning dynamics in which travelers repeatedly play a routing game in a transportation network. Travelers have access to a public information system, which updates and broadcasts the probability of each possible state (public belief). We first describe the routing game and travelers' routing decisions in each stage. Then, we present how the information system updates the belief of the state. 

\subsection{Routing Gams}
Consider a transportation network modeled as a directed graph with a single origin-destination pair. Let $\E$ denote the set of edges and $\R$ denote the set of routes. 
The uncertain network state, denoted $\s$, represents the unknown infrastructure condition after disruption. The cost (travel time) function of each edge $\e$, denoted $\ces(\cdot)$, depends on the state. We assume that the state is grounded in a set $\S$, and fixed throughout the learning dynamics. %

In each stage game $\gamet$, where $\t = 1, 2, \dots$, travelers receive the public belief of the state from the public information system. The belief is denoted as $\thetat = \(\thetat(\s)\)_{\s \in \S} \in \Delta(\S)$, where $\thetat(\s)$ is the probability of state $\s$. 

Travelers are non-atomic players with total demand of $\D$. Travelers' routing strategy is $\qt\deleq\(\qrt\)_{\r \in \R}$, where $\qrt$ is the demand of travelers using route $\r$. A strategy is feasible if it satisfies the following conditions: 
\begin{align*}
\sum_{\r \in \R} \qrt&= \D, \\
\qrt  &\geq 0, \quad \forall \r \in \R.
\end{align*}The set of feasible strategy is denoted by $\Q$. Given any strategy $\qt\in \Q$, the aggregate edge load vector in stage $\t$ is denoted as $\wt=\(\wet\)_{\e \in \E}$, where
\begin{align*}
\wet=\sum_{\r \ni \e} \qrt, \quad \forall \e \in \E.
\end{align*}


The cost on edge $\e$ in stage $\t$, denoted $\cet$, is the sum of a \emph{state-dependent} edge cost function $\ces(\wet)$, and a random variable $\epet$ representing the noise of the realized cost:
\begin{align}\label{cet}
\cet=\ces(\wet)+\epet, \quad \forall \e \in \E,
\end{align}
We make the following assumptions on the edge cost functions and the random noise: 
\begin{itemize}
\item[(A1)] For any $\e \in \E$, the edge cost function $\ces$ is strictly increasing in $\wet$, and the derivative on $\wet$ is bounded from below by a positive number $\alpha>0$, i.e. 
\begin{align*}
\frac{d \ces(\wet)}{d \wet} \geq \alpha, \quad \forall \e \in \E, \quad \forall \s \in \S, \quad \forall \wet>0.
\end{align*}
\item[(A2)] The stage noise $\ept=(\epet)_{\e \in \E}$ has Gaussian distribution with mean $\mathbf{0}$ and a non-degenerate covariance matrix $\Sigma$. The random variables $\left\{\ept\right\}_{\t=1}^{\infty}$ are independently and identically distributed. 
\end{itemize} 
With a slight abuse of notation, we denote the cost function of each route $\r \in \R$ in state $\s$ as $\crs(\qt)= \sum_{\e \in \r} \ces(\wet)$. The realized cost on $\r$ is $\crt=\crs(\qt) +\sum_{\e \in \r} \epet$. Then, one can compute the expected cost of each $\r \in \R$ based on the stage belief $\thetat$ as follows: 
\begin{align*}
\ecrqt=\sum_{\s \in \S} \thetat(\s) \crs(\qt). 
\end{align*}
We assume that travelers are myopic players in that they minimize the expected cost of the chosen route in each stage, without considering the impact of their route choice on the future outcomes. Then, travelers must take routes with the smallest expected cost based on the public belief. This leads to the concept of Wardrop equilibrium defined as follows: 
\begin{definition}\label{def_eq}
A feasible strategy profile $\qtwe \in \Q$ is a Wardrop equilibrium corresponding to the stage belief $\thetat \in \Delta(\S)$ if the following condition is satisfied:
\begin{align*}
\noindent\qrtwe > 0 \text{} \Rightarrow \text{ } \ecrqtwe=\min_{\r^{'}\in \R} \ecrpqtwe, \quad \forall \r \in \R.
\end{align*}
\end{definition}
Since the edge cost functions are increasing in the aggregate edge load, the equilibrium is \emph{essentially unique} in that given any $\thetat$, the corresponding equilibrium edge load $\wtwe=\(\wewet\)_{\e \in \E}$ is unique (see \cite{sandholm2001potential}). 

If $\theta$ assigns probability 1 on state $\s$, then Definition \ref{def_eq} reduces to the classical Wardrop equilibrium under complete information of the state $\s$. We denote the complete information equilibrium edge load vector as $w^{\s*}$. 


\subsection{Information and Belief}\label{sub:belief}
We introduce a public information system, which can observe the stage game outcomes -- the load of each edge and the realized costs on edges that are used in each stage. The information system then updates the public belief based on the stage game outcomes according to Bayes' rule, and broadcasts the updated public belief to all travelers in the next stage. 


The initial public belief, denoted $\thetazero$, represents the information of the network condition that is available to the public information system. We assume that the initial belief does not exclude any possible state: 
\begin{align}\label{continuity}
\centering
\thetazero(\s)>0, \quad \forall \s \in \S.
\end{align}

We now introduce our learning dynamics. The initial time can be set to the time of change in the existing network condition (for example, after a disruption as motivated in Sec. \ref{sec:intro}). In each stage $\t \in \{1, 2, \dots\}$: \\
\emph{Step 1:} The information system broadcasts the public belief of the state $\theta^{\t-1}$ to all travelers.\\
\emph{Step 2:} Traveler populations play the routing game $\gamet$ based on the public belief $\theta^{\t-1}$ according to Wardrop equilibrium $\qtwe$. The induced edge load vector is $\wtwe$. \\
\emph{Step 3}: The information system observes the aggregate load vector $\wt$, and the realized costs on the taken edges $\ct$. We denote the set of edges that are used in stage $\t$ as $\Et \deleq \{\e \in \E|\wtwe_e >0\}$; and thus $\ct=\(\cet\)_{\e \in \Et}$. In state $\s$, the observed costs on these edges has the following Gaussian distribution: 
\begin{align}\label{ct_distribution}
\ct ~ \sim ~ N\(\ell_{\Et}^{\s}, \Sigt\), 
\end{align}
where $\ell_{\Et}^{\s}=\(\ces(\wtwe_e)\)_{\e \in \Et}$, and $\Sigt$ is the sub-matrix of $\Sigma$ with rows and columns corresponding to edges in $\Et$. Then, the probability density function of $\ct$ for a state $\s$ and edge load vector $\wtwe$ can be written as follows: 
 \begin{align}\label{density}
\phi^\t[\s, \wtwe](\ct)=\frac{exp \left\{ -\frac{1}{2} (\ct-\ell_{\Et}^{\s})' \(\Sigt\)^{-1}(\ct-\ell_{\Et}^{\s}) \right\}}{(2\pi)^{\frac{|\Et|}{2}}\sqrt{|\Sigt|}}.
\end{align}  The belief of the state is updated using Bayes' rule:
  \begin{align}\label{Bayesian}
\theta^{\t+1}(\s)= \frac{\theta^{\t}(\s) \cdot \phi^\t[\s, \wtwe](\ct)}{\sum_{\s^{'} \in \S}\theta^{\t}(\s^{'}) \cdot  \phi^\t[\s', \wtwe](\ct)} \quad   \forall \s \in \S.
\end{align}We emphasize that travelers' routing decisions impact the belief update in \eqref{Bayesian} in two respects. Firstly, the induced edge load impacts the distribution of realized costs, and hence the belief update. Secondly, in each stage, if an edge is not used by travelers, then the information system does not observe the cost of that edge. In such cases, travelers are not able to learn how the state impacts the cost on unused edges. 

Note that the distribution of $\ct$ only depends on the state $\s$ and the edge load $\wt$. Then, given any two stages $\t$ and $\t'$, the realized costs $\ct$ and $c^{\t'}$ are independent conditional on the edge loads $\wt$ and $w^{\t'}$. We define $\whis^{\t*} \deleq \(w^{j*}\)_{j=1}^{\t}$ as the history of equilibrium edge load vectors, and $\playpath^{\t} \deleq \(c^j\)_{j=1}^{\t}$ as the history of realized costs until the end of stage $\t$. Then in state $\s$, the probability density function of $\playpatht$ conditioned on $\whis^{\t*}$, can be written as follows: 
\begin{align}\label{phi_fun}
\ps\(\playpatht|\whis^{\t*}\)&=\prod_{j=1}^{\t}\phi^j[\s, w^{j*}](c^j),  \quad \forall \t.
\end{align}
By iteratively applying \eqref{Bayesian}, we can derive $\thetat$ from the initial belief $\thetazero$ as follows:
  \begin{align}\label{update_playpath}
\thetat(\s)=&\frac{\thetazero(\s) \cdot p^{\s}(\playpath^{\t}|\whis^{\t*})}{\sum_{\s' \in \S} \thetazero(\s') \cdot p^{s'}(\playpath^{\t}|\whis^{\t*})}, \quad \forall \s \in \S, \quad \forall \t 
\end{align}It is clear that the main advantages of introducing the information system in our model is that it observes the stage game outcomes, and computes belief updates. This setup has two advantages: First is that travelers do not need to observe the outcomes and update beliefs by themselves. Second is that the set of travelers participating in each stage game can be different, i.e. an individual traveler may participate in one or multiple routing games as long as they are informed of the public belief of the state in the participated stages. 

In our learning model, the public belief of the state $\thetat$ changes based on the stage game outcomes, and in turn influences the edge load $\wtwe$ induced by travelers' route choices. Therefore, the tuple $(\thetat, \wtwe)$ governs how travelers learn the state of the network and how they make route choices according to our learning dynamics.  

\section{Convergence of Learning Dynamics}\label{sec_self_confirm}
In this section, we show that the sequence $\{(\thetat, \wtwe)\}_{\t=1}^{\infty}$ generated by our learning dynamics converges to a rest point with probability 1. 

Before presenting our theorem, we first introduce the concept of distinguishable states. To avoid confusion, we denote the true state as $\sran$. For any load vector $w$, we say that the state $\s$ is \emph{distinguishable} from the true state $\sran$ if: 
\begin{align*}
\exists~\e \in \{\E|\we>0\}, \quad s.t. \quad \ces(\we) \neq\cesran(\we). 
\end{align*}
The set of distinguishable states given $\w$ is denoted as $\S^{\dagger}(\w)$. The distribution of the realized costs in any distinguishable state $\s \in \S^{\dagger}(\w)$ as in \eqref{ct_distribution} is different from that in the true state $\sran$. Hence, state $\s$ can be distinguished from the true state $\sran$ based on the realized costs. However, if the cost functions in state $\s$ are different from that in $\sran$ only on a subset of edges, and no edge in that set is used in $\w$, then $\s$ is indistinguishable, i.e. $\s \in \S \setminus \S^{\dagger}(\w)$. 

\begin{theorem}\label{theorem_convergence}
For any true state $\sran \in \S$, and any initial belief $\thetazero$ that satisfies \eqref{continuity}, we have: 
\begin{subequations}
\begin{align}
&\lim_{\t \to \infty} \thetat = \thetabar, \quad \text{w.p.1}, \label{belief_converge}\\
&\lim_{\t \to \infty} \wtwe = \wbar, \quad \text{w.p.1}, \label{strategy_converge}
\end{align}
\end{subequations}
where $\thetabar$ is a probability vector that satisfies:
 \begin{align}\label{eq:confirm}
\thetabar(\s)=0, \quad  \forall \s \in \S^{\dagger}(\wbar).  
\end{align}
and $\wbar$ the equilibrium edge load vector corresponding to the public belief $\thetabar$.
\end{theorem}
From Theorem \ref{theorem_convergence}, we know that the public belief $\thetat$ and the associated equilibrium edge load $\wtwe$ in the learning dynamics eventually converge to a rest point $\(\thetabar, \wbar\)$, which satisfies the following two properties: 
\begin{enumerate}
\item \emph{Equilibrium under imperfect state information:} Travelers have no incentive to deviate from the chosen routes based on public information of the state. 
\item \emph{Consistency:}
Since the public belief $\thetabar$ excludes any distinguishable states, for any $\s$ such that $\thetabar(\s)>0$ and any $\e \in \Ebar$, we must have $\ces(\wbar_e)=\cesran(\wbar_e)$. Then, travelers accurately learn the cost of edges that are used, i.e. $\mathbb{E}_{\theta}[\ces(\wbar)]=\ell_\s^\sran(\wbar)$ for any $\e \in \Ebar$. 
\end{enumerate}

Additionally, we can show that the convergent public belief $\thetabar$ is a fixed point of the belief update function in \eqref{Bayesian}. If the initial belief $\thetazero=\thetabar$, then in any stage $\t$, the public belief is $\thetat=\thetabar$, and the equilibrium edge load is $\wtwe=\wbar$.

We present the main intuition behind the proof of Theorem \ref{theorem_convergence}. 

Firstly, we show that the process of the public beliefs $\(\thetat\)_{\t=0}^{\infty}$ is a bounded martingale. Hence, $\thetat$ converges to a random probability vector $\thetabar$ with probability 1. 

Secondly, by adapting the sensitivity analysis approach in \cite{dafermos1984sensitivity} to our problem, we show that the equilibrium edge load $\wtwe$ in each stage is a continuous function of the public belief $\theta^{\t-1}$. Therefore, from continuous mapping theorem and the fact that $\thetat$ converges to $\thetabar$, we can show that the unique equilibrium edge load $\wtwe$ also converges to $\wbar$, which is the equilibrium edge load corresponding to $\thetabar$. 

Finally, it remains to be shown that $\thetabar$ satisfies \eqref{eq:confirm} with probability 1. To start with, we view the problem of distinguishing any state $\s$ from the true state $\sran$ as a hypothesis testing problem. Next, we can analyze the log-likelihood ratio of the sequence of realized cost conditional on the edge loads until stage $\t$ in state $\s$ and the true state $\sran$, denoted $\log\(\frac{p^{\sran}(\playpath^{\t}|W^{\t*})}{\ps(\playpath^{\t}|W^{\t*})}\)$. In fact, wecan conclude that for any distinguishable state $\s \in \S^{\dagger}(\wbar)$, $\lim_{\t \to \infty} \log\(\frac{p^{\sran}(\playpath^{\t}|W^{\t*})}{\ps(\playpath^{\t}|W^{\t*})}\)=\infty$ with probability 1. 
Since the initial public belief satisfies $\thetazero(\sran)>0$, 
\begin{align*}
&\lim_{\t \to \infty}\thetat(\s) \stackrel{\eqref{update_playpath}}{=}\lim_{\t \to \infty}\frac{\thetazero(\s) \cdot \ps(\playpath^{\t}|W^{\t*})}{\sum_{\s' \in \S} \thetazero(\s') \cdot p^{s'}(\playpath^{\t}|W^{\t*})}\leq \lim_{\t \to \infty}\frac{\thetazero(\s) \cdot \ps(\playpath^{\t}|W^{\t*})}{\thetazero(\s) \cdot \ps(\playpath^{\t}|W^{\t*})+ \thetazero(\sran) \cdot p^{\sran}(\playpath^{\t}|W^{\t*})}\\
=& \lim_{\t \to \infty}\frac{\thetazero(\s) }{\thetazero(\s) + \thetazero(\sran) \cdot \frac{p^{\sran}(\playpath^{\t}|W^{\t*})}{\ps(\playpath^{\t}|W^{\t*})}}=0, \quad \forall \s \in \S^{\dagger}(\wbar).
\end{align*}Hence, $\thetabar$ satisfies \eqref{eq:confirm}.

\section{Rest points analysis}\label{sec:rest_point}
In this section, we first compare travelers' route choices and average cost at a rest point with that in complete information equilibrium. Then, we provide conditions under which travelers eventually make route choice as if they know the true state with probability 1. Finally, we illustrate some interesting aspects of learning dynamics and rest points through examples. 

\subsection{Compaison with Complete Information Equilibium}
We say the learning is \emph{complete} if the edge load at the rest point is $\wbar= \wsran$, which is the complete information equilibrium in the true state $\sran$, with probability 1. 

However, learning may converge to other rest points $\(\thetabar, \wbar\)$ such that $\wbar\neq\wsran$. This is because $\thetabar$ may assign positive probability on another indistinguishable state $\s\in \S \setminus \S^{\dagger}(\wbar)$ that is not distinguishable from the true state $\sran$ given the convergent load vector $\wbar$. Therefore, although the estimated costs on the used edges are identical to the true costs as, the estimated cost on an unused edge based on $\thetabar$ may be higher than the actual cost in the true state. Consequently, if travelers know the true state, they may have the incentive to deviate from the rest point. 

We denote the average cost of travelers at a rest point with edge load $\wbar$ as $C(\wbar)\deleq \sum_{\e \in \E}\wbar_e\ell_e^\sran(\wbar_e)$, and the averge cost in the complete information equilibrium as $C(\wsran) \deleq \sum_{\e \in \E}\wsran_e\ell_e^\sran(\wsran_e)$. 

The next proposition shows that if the network is series-parallel (i.e. the network does not have an embedded wheatstone network, see \cite{milchtaich2006network}), the average cost at any rest point is no less than that in complete information equilibrium. 
\begin{proposition}\label{higher_cost}
If the network is series-parallel, then $C(\wbar) \geq C(\wsran)$ at any rest point $(\thetabar, \wbar)$.
\end{proposition}
The idea of the proof is that the edge load at any rest point is equivalent to the complete information equilibrium in a routing game on a subnetwork. In other words, travelers make route choice as if they only know a subset of the available routes in the original network. Then, based on Theorem 1 in \cite{milchtaich2006network}, we can show that if the network is series-parallel, the equilibrium average cost on the subnetwork is no less than that on the original network. 

\subsection{Complete Learning}
We now provide a set of conditions, under each of which the learning is complete with probability 1. 

\begin{proposition}\label{nash_proposition}
For any true state $\sran \in \S$, the learning is complete, i.e. $\lim_{\t \to \infty} \wtwe= \wsran$ with probability 1, if any of the following conditions is satisfied: 
\begin{itemize}
\item[(1)] \emph{Fully distinguishable states:} For any $\w$ and any $\s \in \S\setminus \{\sran\}$, state $\s$ is distinguishable from $\sran$.
\item[(2)] \emph{State-independent free flow travel time:} For any $\e \in \E$ and any $\s \in \S$, $\ell_\e^{\s}(0)$ is identical across states.  
\item[(3)] \emph{All edges are utilized:} For any $\s \in \S$ and any $\e \in \E$, $\w_e^{\s*}>0$. \hspace*{\fill} 
\end{itemize}
\end{proposition}

Each condition in (1) - (3) ensures that travelers repeatedly use the set of edges that should be taken in complete information equilibrium. Then, travelers will eventually learn the costs on these edges, and choose routes as if they know the true state. 

In practice, condition (1) is relevant if every state impacts the costs on all edges. Condition (2) requires that the state only impacts the costs when there is congestion ($w_e>0$). For example, lane closure does not change the cost when there is no traffic, but significantly aggrevates congestion due to the loss of capacity. Condition (3) requires that all edges are utilized regardless of the state. This will hold when the traffic demand is high so that travelers must take all routes. 

\subsection{Examples and Discussion}
\begin{figure}[htp]
\centering
\includegraphics[width=0.45\textwidth]{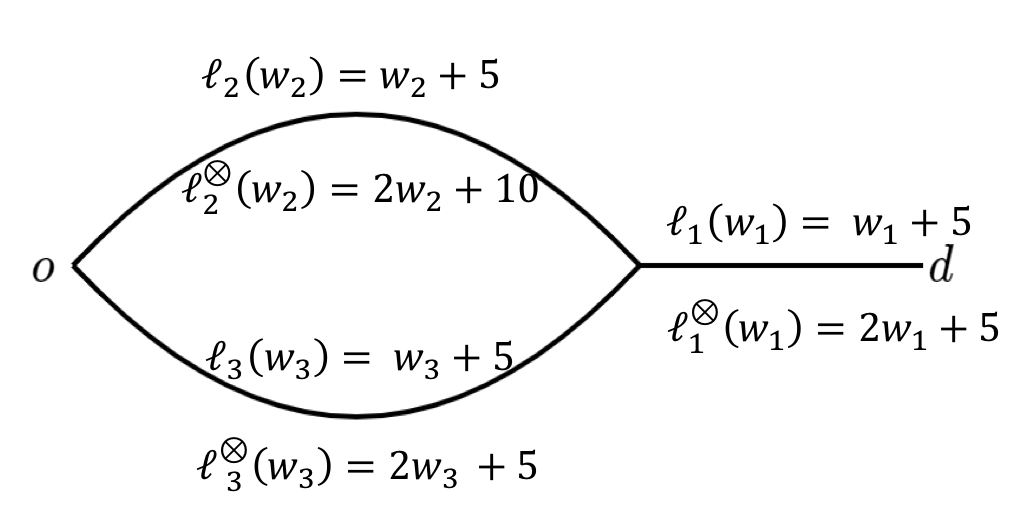}
\caption{Three-edge network}
\label{network}
\end{figure}

Consider the three-edge series-parallel network with the set of states: $\{e_1, e_2, e_3, \emptyset\}$, where $\s=\e$ is the state in which edge $\e$ is compromised, and $\s=\emptyset$ is the state in which no edge is compromised. The cost of edge $\e$ is $\ell_e(w_e)$ if $\s \neq \e$, and $\ell_\e^{\otimes}(\we)$ if $\s=\e$. See Fig. \ref{network} for the network and cost functions. In this example, we assume that the noise term in each stage is $\epsilon^\t=\(\epsilon_\e^\t\)_{\e \in \E} \sim \mathcal{N}\(0, \Sigma\)$, where $\Sigma$ is a three-dimensional identity matrix. The total demand is 1. 

Let the true state be $\sran = \emptyset$. The set of rest points is as follows:
\begin{align*}
\left\{\begin{array}{l}
\theta^{\sran}=(0, 0, 0, 1),\\
\wsran=\(1, 0.5, 0.5\)
\end{array}\right\} \cup \left\{\begin{array}{l}
\thetabar=(0, x, 0, 1-x) \text{ } s.t. \text{ } x\geq 0.2,\\
\wbar=\(1, 0, 1\)
\end{array}\right\}
\end{align*}That is, apart from the complete equilibrium edge load $\wsran$, travelers may exclusively choose the route $e_3$ - $e_1$ if they  believe that the probability of $\s=e_2$ is no less than $0.2$. In this case, the public information system cannot distinguish state $s=e_2$ from the true state $\sran=\emptyset$ based on the realized costs. We can check that $C(\wsran)= 11.5< C(\wbar)=12$.

\begin{figure}[htp]
    \begin{subfigure}[b]{0.36\textwidth}
        \includegraphics[width=\textwidth]{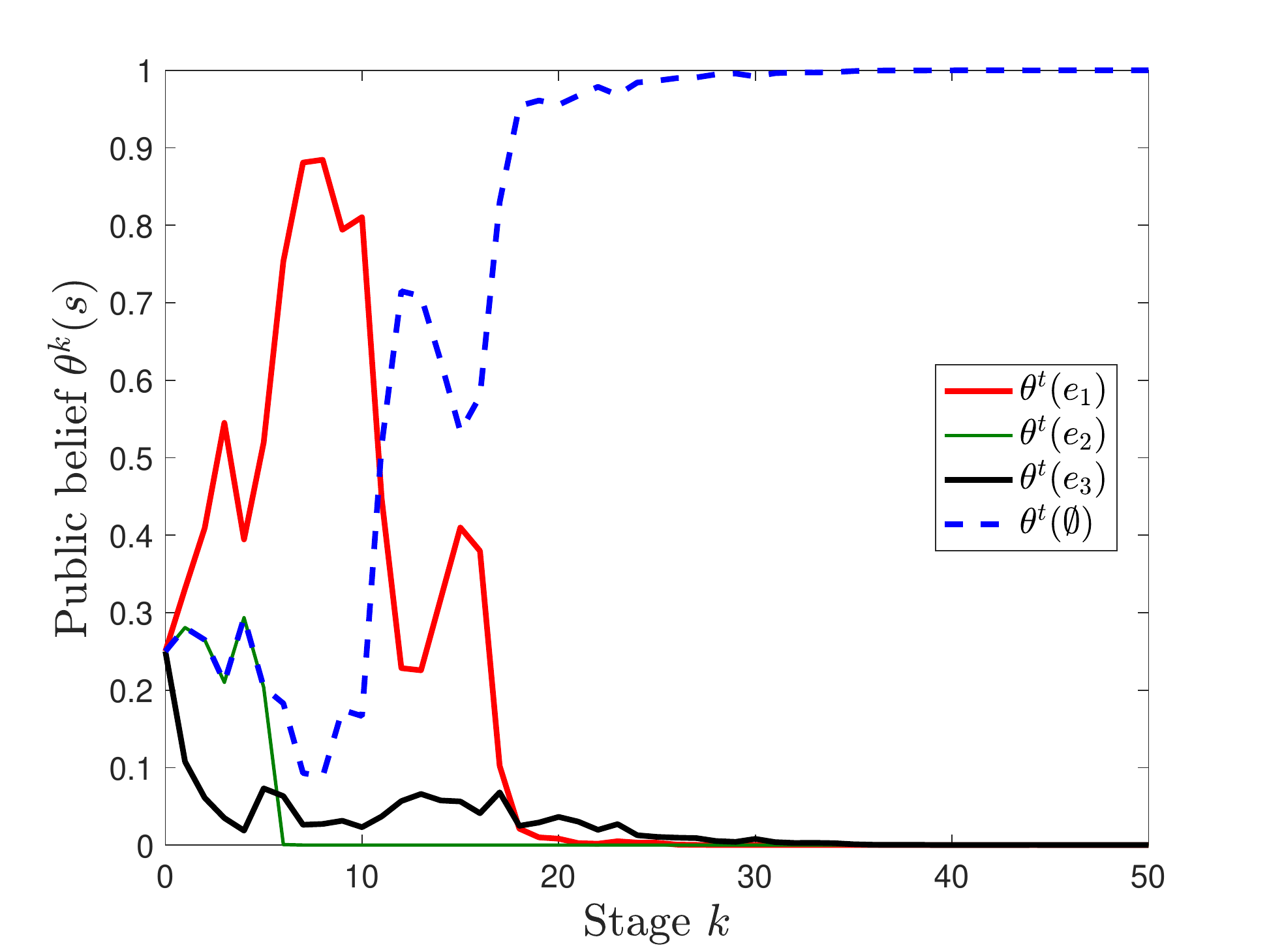}
        \caption{ }
        \label{fig:theta_complete}
    \end{subfigure}
~
\centering
	\begin{subfigure}[b]{0.36\textwidth}
        \includegraphics[width=\textwidth]{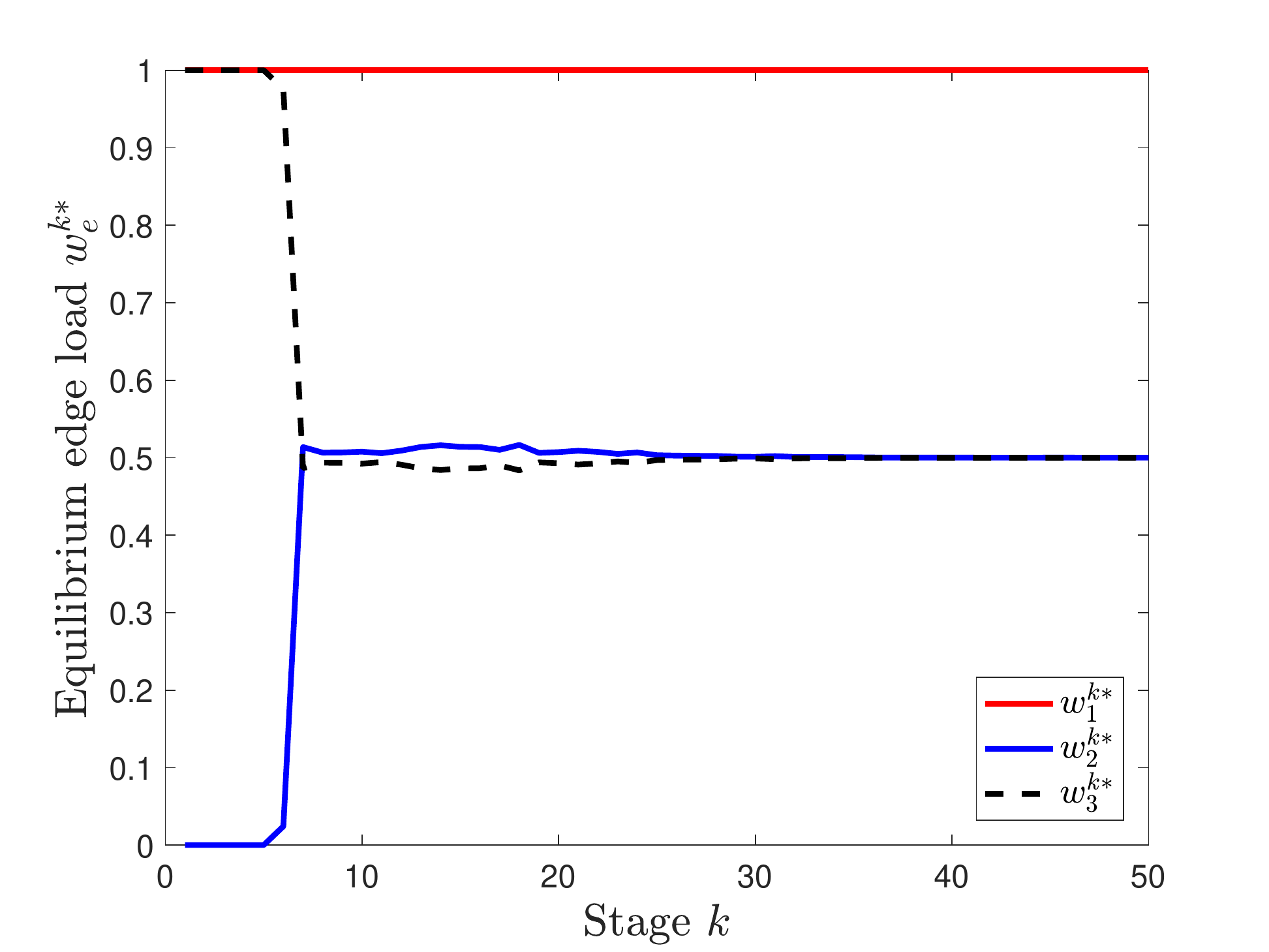}
        \caption{ }
        \label{fig:w_complete}
    \end{subfigure}\\
    \begin{subfigure}[b]{0.36 \textwidth}
        \includegraphics[width=\textwidth]{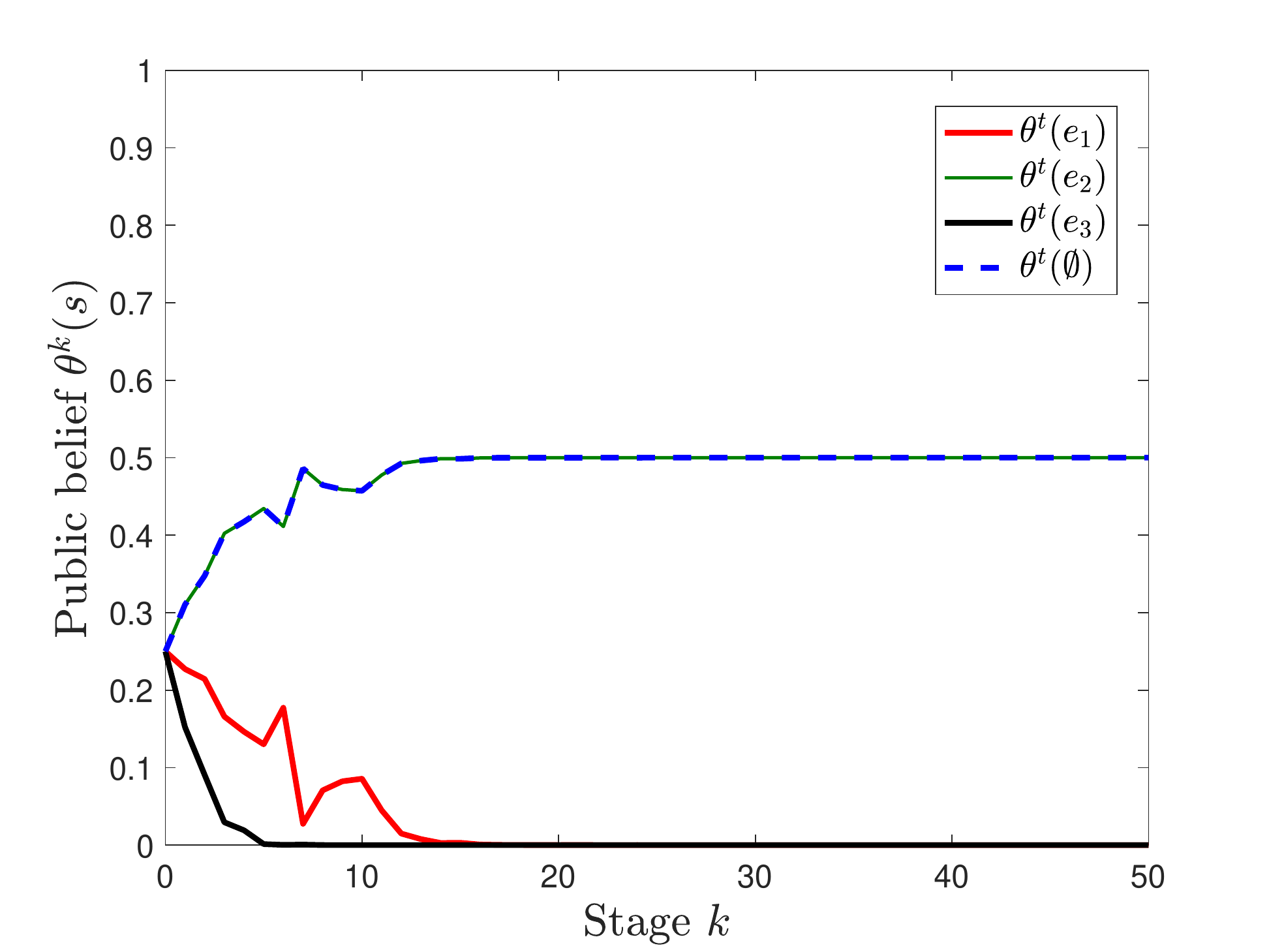}
        \caption{}
        \label{fig:theta_self_confirm}
    \end{subfigure}
~
\centering
	\begin{subfigure}[b]{0.36\textwidth}
        \includegraphics[width=\textwidth]{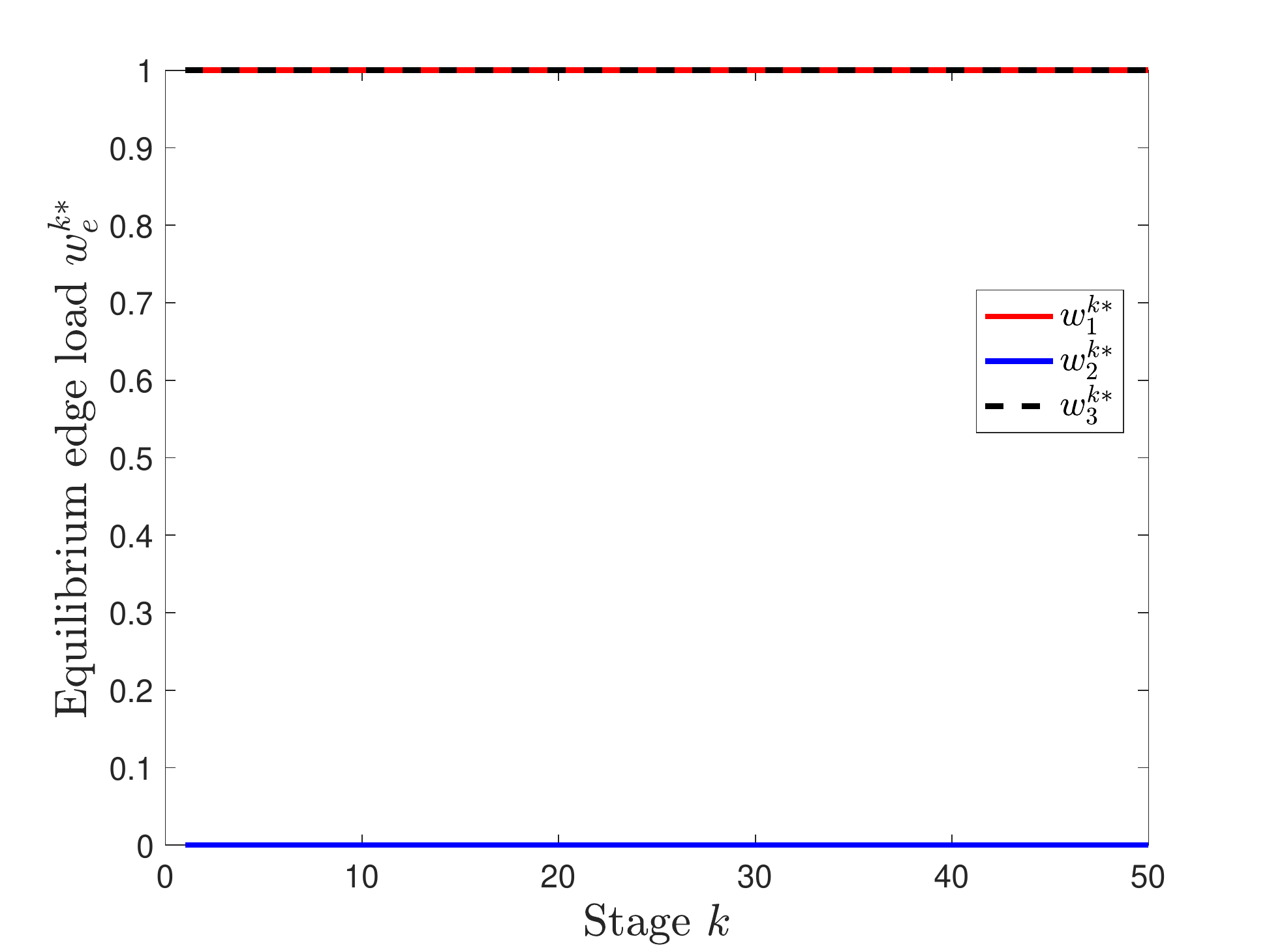}
        \caption{}
        \label{fig:w_self_confirm}
    \end{subfigure}\\
 \begin{subfigure}[b]{0.36 \textwidth}
        \includegraphics[width=\textwidth]{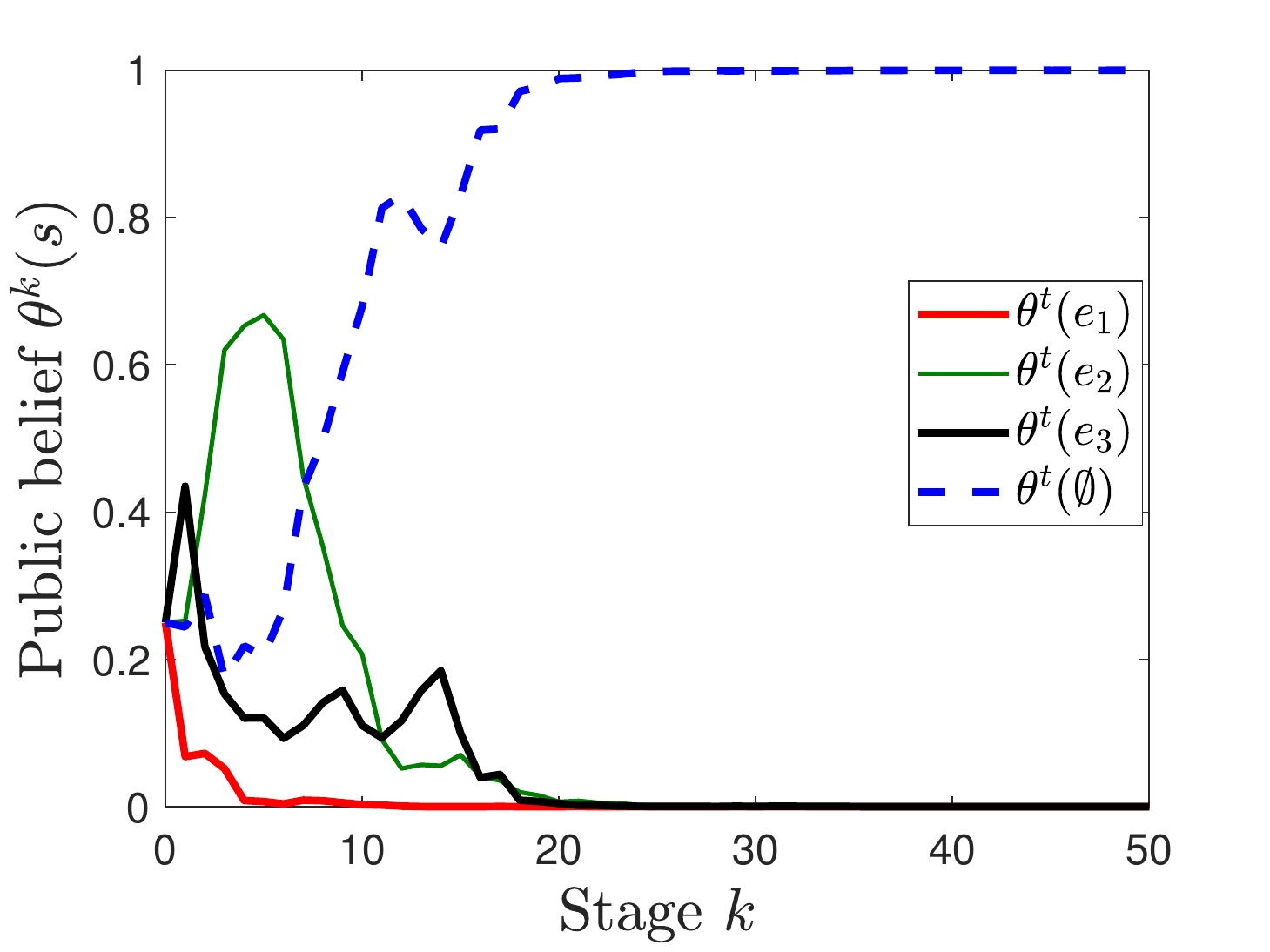}
        \caption{}
        \label{fig:theta_complete_identical}
    \end{subfigure}
~
\centering
	\begin{subfigure}[b]{0.36\textwidth}
        \includegraphics[width=\textwidth]{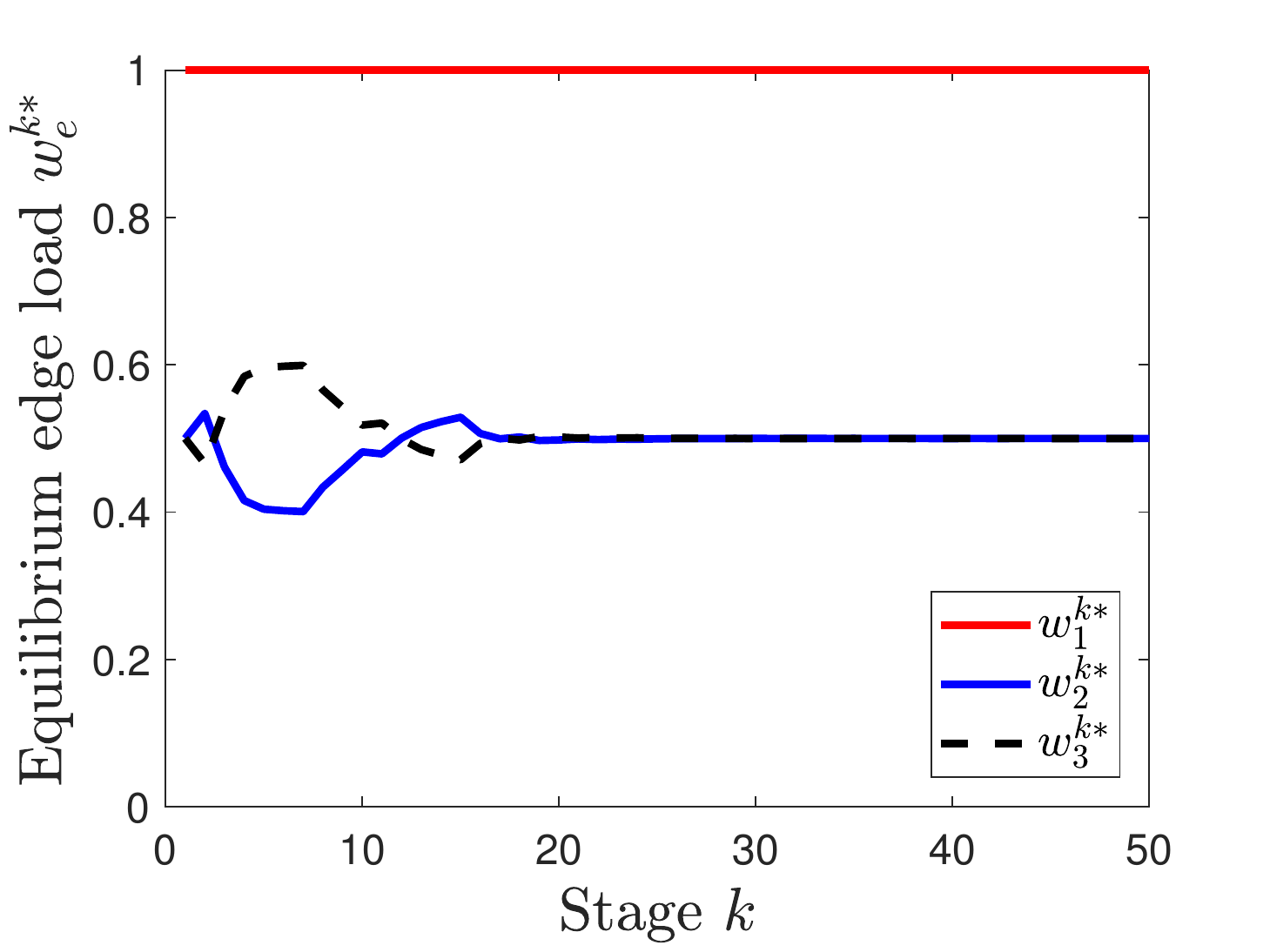}
        \caption{}
        \label{fig:w_complete_identical}
    \end{subfigure}\\
 \begin{subfigure}[b]{0.36 \textwidth}
        \includegraphics[width=\textwidth]{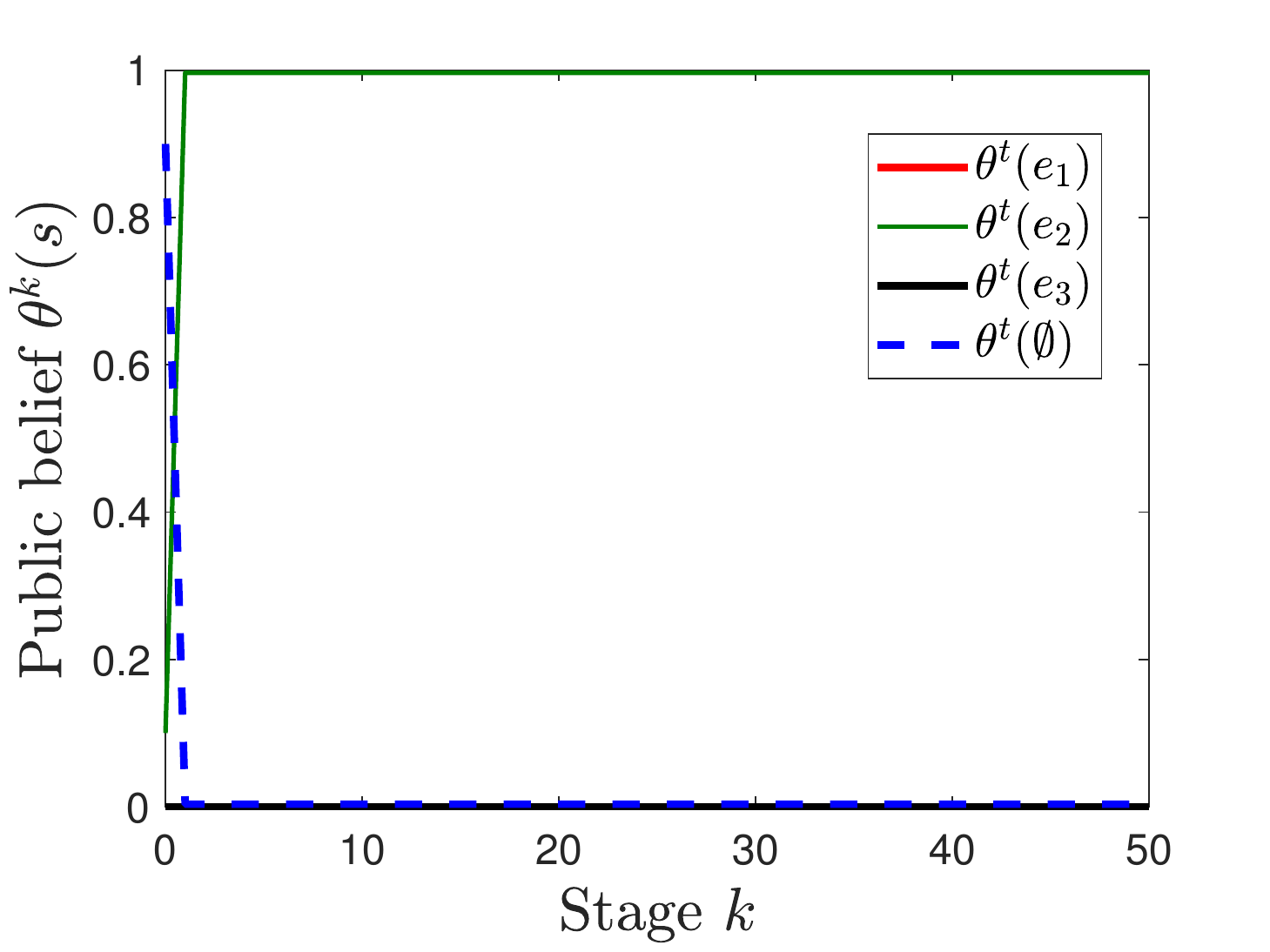}
        \caption{}
        \label{fig:theta_shock}
    \end{subfigure}
~
\centering
	\begin{subfigure}[b]{0.36\textwidth}
        \includegraphics[width=\textwidth]{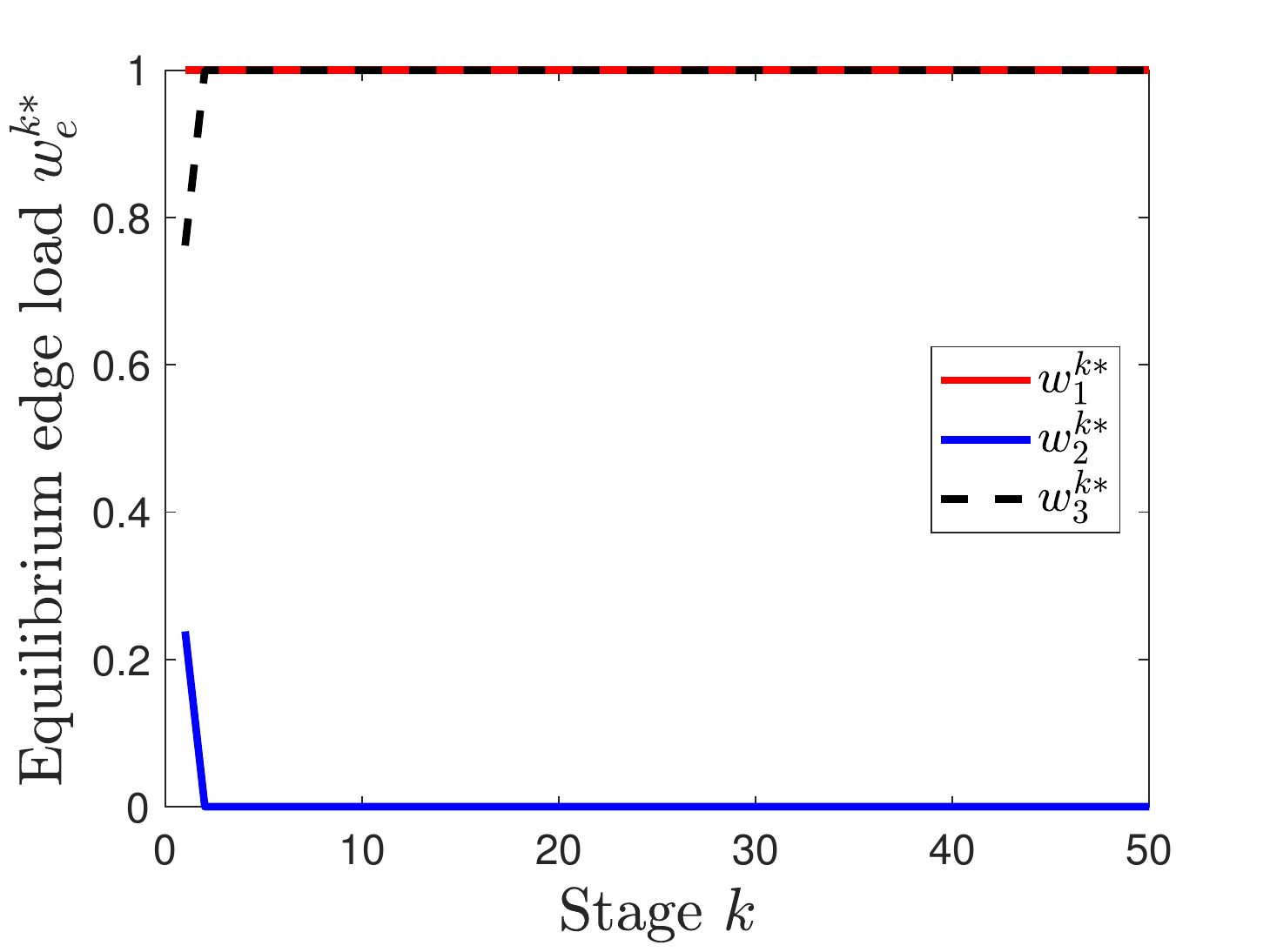}
        \caption{}
        \label{fig:w_shock}
    \end{subfigure}
    \caption{Public belief and equilibrium edge load in learning dynamics: (a) - (b) Learning is complete; (c) - (d) Learning is incomplete; (e) - (f) State-independent free-flow travel time ensures complete learning; (g) - (h) Noisy cost realizations lead to incomplete learning.}
\label{fig:learning_demonstrate}
\end{figure}

We simulate the learning dynamics with the initial public belief $\thetazero = \(1/4, 1/4, 1/4, 1/4\)$, which satisfies \eqref{continuity}. Figures \ref{fig:theta_complete} - \ref{fig:w_complete} demonstrate the public belief $\thetat$ and the equilibrium edge load $\wtwe$ in each stage for a play-path that converges to $(\theta^{\sran}, \wsran)$, i.e. learning is complete. Figures \ref{fig:theta_self_confirm} - \ref{fig:w_self_confirm} illustrate a play-path that converges to $(\thetabar, \wbar)$, where $\thetabar=\(0, 1/2, 0, 1/2\)$, i.e. learning is not complete. 

Moreover, if the edge cost on $e_2$ when it is compromised is $\ell_2^{\otimes}=2w_2+5$. Then, the cost functions satisfy the condition (2) in Proposition \ref{nash_proposition}. We can check from Fig. \ref{fig:theta_complete_identical} - \ref{fig:w_complete_identical} that the learning is complete. 

Finally, we assume that learning starts with $\thetazero=\(0, 0.1, 0, 0.9\)$.
Fig. \ref{fig:theta_shock} - \ref{fig:w_shock} gives a play-path, in which learning is not complete even if the the initial public belief is $90\%$ accurate. Although travelers takes $e_2$ in the first stage, the cost realized from Normal distribution happens to be very high. The information system is not able to tell whether the high cost is due to compromised facility or due to random noise in that stage. Consequently, the updated belief assigns high probability on state $\s=e_2$, and travelers no longer take $e_2$ in future stages. 

Our results and examples show that to ensure complete learning in general, apart from the information obtained from repeated routing, complementary approaches are needed to measure the costs of edges that are not taken by myopic travelers. One direction to extend our work is to analyze how to efficiently estimate the costs on these edges by placing sensors, sending out probing teams or incentivizing explorations. Note that deployment of these measurement resources should account for travelers' strategic route decisions, the level of noise in realized costs, as well as how the information will impact the route choices and costs at rest points.

\section{Concluding Remarks}\label{sec:conclude}
In this article, we study how strategic travelers learn the uncertain state after infrastructure disruptions and adjust their route choices dynamically with the access of a public information system. Our results include the convergence of belief and edge loads, comparison of rest points with complete information equilibrium, and conditions that guarantee complete learning. 


All our results hold for networks with multiple origin-destination pairs. Additionally, future work entails relaxing these assumptions: First, permiting the noise terms to be realized from a variety of distributions that depend on the state and edge load. Second, relaxing the assumption that the total demand is inelastic to incorporate random fluctuation of traffic demand on a daily basis. 

Another future direction of interest is to analyze the learning dynamics in asymmetric and incomplete information environment. In practice, travelers may obtain private information of the state after infrastructure disruption, or have private perception of the belief provided by the public information system. Then, travelers' route choices depend on the heterogeneous private information (perception) as well as the public belief, and the update of public belief in turn depends on the route choices.  Addressing this problem would involve analyzing how the private information impacts the edges that are used in each stage as well as the residual information heterogeneity at the rest points. 

\section*{Acknowledgement}
We are grateful to Prof. Asu Ozdaglar and Prof. Muhamet Yildiz for useful discussions. 
The first author was supported by the IDSS Hammer Fellowship,
and the second author was party supported by NSF CAREER Award.
\bibliographystyle{plainnat}
\bibliography{ifacconf}

\end{document}